\documentclass[a4paper,twocolumn,showpacs,amsmath,amssymb,floatfix,prl,preprintnumbers,footinbib]{revtex4}
\usepackage{graphicx}
\usepackage{dcolumn}
\usepackage{bm}

\begin{document}


\title{Comment: Energy Spectrum of a Graphene Quantum Dot in a Perpendicular Magnetic Field}
\author{S.~Schnez$^1$, K.~Ensslin$^1$, M.~Sigrist$^2$, and T.~Ihn$^1$}

\affiliation{$^1$Solid State Physics Laboratory, ETH Z\"urich, 8093 Z\"urich, Switzerland\\
$^2$Institute for Theoretical Physics, ETH Z\"urich, 8093 Z\"urich, Switzerland}

\begin{abstract}
In a recent comment \cite{Falaye16}, Falaye {\it et al.} claim that there are certain flaws in our publication \cite{Schnez08}. We point out that our results, in particular the analytic derivation of the energy spectrum of a circular graphene quantum dot exposed to a perpendicular magnetic field, are correct and equivalent to the result of Falaye {\it et al.}. A misleading notation error is corrected.
\end{abstract}

\maketitle



Falaye {\it et al.} claim \cite{Falaye16} that there are certain flaws in our publication \cite{Schnez08}, in particular that the wave functions given by Eq.~5 in Ref.~\cite{Schnez08} cannot be normalized and that, correspondingly, the implicit equation Eq.~6 describing the energy spectrum is incorrect. We note the following:
\begin{itemize}
	\item The mathematical derivation based on our {\it ansatz} as described in Ref.~\cite{Schnez08} is correct. As a matter of fact, the results of Falaye {\it et al.}, who use the confluent hypergeometric function instead of the generalized Laguerre polynomials, are equivalent to our results. The parameter $a$ in the generalized Laguerre polynomials $L(a,b,x)$ can take real values, not only integers as in Ref.~\cite{Falaye16}. This is beyond the definition in Ref.~\cite{Abramowitz}, but well-defined and used today (also implemented in e.g. Mathematica).
	\item Our definition of the quantum number $n$ differs from the definition in Ref.~\cite{Falaye16}. They do not denote the same quantity.
	\item Using a recursion theorem for the generalized Laguerre polynomials \cite{Abramowitz}, the energy spectrum Eq.~6 in Ref.~\cite{Schnez08} can be written in a more compact form as (as pointed out by Falaye {\it et al.})
	\begin{equation}\label{spectrum}
		\begin{split}
			&L\left(\frac{k^2 l_B^2}{2}-m-1, m+1, \frac{R^2}{2l_B^2}\right) \\
			&- \tau \frac{kl_B}{R/l_B}\cdot L\left(\frac{k^2 l_B^2}{2}-m-1, m, \frac{R^2}{2l_B^2}\right) = 0.
		\end{split}
	\end{equation} 
\end{itemize} 

The use of the parameter $m$ in Eq.~11 of our publication \cite{Schnez08} is incorrect. Rather, it should read
\begin{equation}
	E = \pm v_F \sqrt{2e\hbar B(m+1+p)},
\end{equation}
where $m$ is the previously defined quantum number and $p$ is an integer with $p > -(m+1)$. This follows from the fact that Eq.~6 in Ref.~\cite{Schnez08} or Eq.~\ref{spectrum} above, respectively, can be simplified to $\left(\Gamma(\alpha)\Gamma(-\alpha)\right)^{-1}=0$ in the limit $R/l_B \to \infty$ with $\alpha:= k^2l_B^2/2-m-1$. This is fulfilled for $\alpha=\pm p$ and $p$ being an integer. The later restriction of $p>-(m+)$ is then required to make the radicand non-negative.

\end{document}